\documentclass[12pt]{article}
\usepackage{psfrag}
\usepackage{graphicx}
\usepackage{latexsym,amsfonts}
\usepackage{amssymb}
\begin{document}
\setcounter{page}{1}
\def\theequation{\arabic{section}.\arabic{equation}}
\def\theequation{\thesection.\arabic{equation}}
\setcounter{section}{0}

\title{On kaonic hydrogen \\ Phenomenological quantum field theoretic
  model revisited}

\author{A. N. Ivanov\,\thanks{E--mail: ivanov@kph.tuwien.ac.at, Tel.:
+43--1--58801--14261, Fax: +43--1--58801--14299}~\thanks{Permanent
Address: State Polytechnic University, Department of Nuclear Physics,
195251 St. Petersburg, Russian Federation}\,,
M. Cargnelli\,\thanks{E--mail: michael.cargnelli@oeaw.ac.at}\,,
M. Faber\,\thanks{E--mail: faber@kph.tuwien.ac.at, Tel.:
+43--1--58801--14261, Fax: +43--1--58801--14299}\,, H.
Fuhrmann\,\thanks{E--mail: hermann.fuhrmann@oeaw.ac.at}\,,\\
V. A. Ivanova\,\thanks{E--mail: viola@kph.tuwien.ac.at, State
Polytechnic University, Department of Nuclear Physics, 195251
St. Petersburg, Russian Federation}\,, J. Marton\,\thanks{E--mail:
johann.marton@oeaw.ac.at}\,, N. I. Troitskaya\,\thanks{State
Polytechnic University, Department of Nuclear Physics, 195251
St. Petersburg, Russian Federation}~~, J. Zmeskal\,\thanks{E--mail:
johann.zmeskal@oeaw.ac.at}}

\date{\today}

\maketitle
\vspace{-0.5in}

\begin{center}
{\it Atominstitut der \"Osterreichischen Universit\"aten,
Arbeitsbereich Kernphysik und Nukleare Astrophysik, Technische
Universit\"at Wien,  Wiedner Hauptstrasse 8-10, \\A-1040 Wien,
\"Osterreich and \\ Stefan Meyer Institut f\"ur subatomare Physik, 
\"Osterreichische Akademie der Wissenschaften,
Boltzmanngasse 3, A-1090, Wien, \"Osterreich} 
\end{center}

\begin{center}

\begin{abstract}
  We argue that due to isospin and $U$--spin invariance of strong
  low--energy interactions the S--wave scattering lengths $a^0_0$ and
  $a^1_0$ of $\bar{K}N$ scattering with isospin $I=0$ and $I = 1$
  satisfy the low--energy theorem $a^0_0 + 3 a^1_0 = 0$ valid to
  leading order in chiral expansion. In the model of strong
  low--energy $\bar{K}N$ interactions at threshold (EPJA {\bf 21},11
  (2004)) we revisit the contribution of the $\Sigma(1750)$ resonance,
  which does not saturate the low--energy theorem $a^0_0 + 3 a^1_0 =
  0$, and replace it by the baryon  background with properties
  of an $SU(3)$ octet. We calculate the S--wave scattering amplitudes
  of $K^-N$ and $K^-d$ scattering at threshold.  We calculate the
  energy level displacements of the ground states of kaonic hydrogen
  and deuterium.  The result obtained for kaonic hydrogen
  agrees well with recent experimental data by the DEAR Collaboration.
  We analyse the cross sections for elastic and inelastic $K^-p$
  scattering for laboratory momenta $70\,{\rm MeV}/c < p_{_K} < 150\,{\rm MeV}/c$ of the incident $K^-$ meson.  The
  theoretical results agree with the available experimental data
  within two standard deviations.
\end{abstract}


\end{center}

\newpage

\section{Introduction}
\setcounter{equation}{0}

Recently in Ref.\cite{IV3} we have proposed a phenomenological quantum
field theoretic model for strong low--energy $K^-p$ interactions at
threshold for the analysis of the experimental data by the DEAR
Collaboration Refs.\cite{DEAR,SIDDHARTA} on the energy level
displacement of the ground state of kaonic hydrogen
\begin{eqnarray}\label{label1.1}
  - \epsilon^{(\exp)}_{1s} + i\,\frac{\Gamma^{(\exp)}_{1s}}{2} &=& 
  (- 194 \pm 37\,(\rm stat.) \pm 6\,(\rm syst.))\nonumber\\
 &+& i\,(125 \pm 56\,(\rm stat.) 
\pm 15\,(\rm syst.))\,{\rm eV}. 
\end{eqnarray} 
According to the Deser--Goldberger--Baumann--Thirring--Trueman formula
(the DGBTT) Ref.\cite{SD54}, the energy level displacement of the ground
state of kaonic hydrogen is related to the S--wave amplitude
$f^{K^-p}_0(0)$ of $K^-p$ scattering at threshold as
\begin{eqnarray}\label{label1.2}
  - \,\epsilon_{1s} + i\,\frac{\Gamma_{1s}}{2} &=& 
  2\,\alpha^3\mu^2 
  \,f^{K^-p}_0(0) =  412.13\,f^{K^-p}_0(0),
\end{eqnarray}
where $\mu = m_K m_p/(m_K + m_p) = 323.48\,{\rm MeV}$ is the reduced
mass of the $K^-p$ pair, calculated for $m_K = 493.68\,{\rm MeV}$ and
$m_p = 938.27\,{\rm MeV}$, and $\alpha = 1/137.036$ is the
fine--structure constant Ref.\cite{PDG04}.  The theoretical accuracy
of the DGBTT formula Eq.(\ref{label1.2}) is about $3\,\%$ including
the vacuum polarisation correction Ref.\cite{TE04}.

For a non--zero relative momentum $Q$ the amplitude $f^{K^-p}_0(Q)$ is
defined by
\begin{eqnarray}\label{label1.3}
f^{K^-p}_0(Q) = \frac{1}{2iQ}\,\Big(\eta^{K^-p}_0(Q)\,
e^{\textstyle\,2i\delta^{K^-p}_0(Q)} - 1\Big),
\end{eqnarray}
where $\eta^{K^-p}_0(Q)$ and $\delta^{K^-p}_0(Q)$ are the inelasticity
and the phase shift of the reaction $K^- + p \to K^- + p$,
respectively.  

The real part ${\cal R}e\,f^{K^-p}_0(0)$ of $f^{K^-p}_0(0)$ defines
the S--wave scattering length $a^{K^-p}_0$ of $K^-p$ scattering
\begin{eqnarray}\label{label1.4}
{\cal R}e\,f^{K^-p}_0(0) = a^{K^-p}_0 = \frac{1}{2}\,(a^0_0 +
a^1_0),
\end{eqnarray}
where $a^0_0$ and $a^1_0$ are the S--wave scattering lengths $a^I_0$
with isospin $I = 0$ and $I = 1$, respectively. 

The imaginary part ${\cal I}m\,f^{K^-p}_0(0)$ of $f^{K^-p}_0(0)$ is
caused by inelastic channels $K^-p \to Y\pi$, where $Y\pi =
\Sigma^-\pi^+$, $\Sigma^+\pi^-$, $\Sigma^0\pi^0$ and
$\Lambda^0\pi^0$, allowed kinematically at threshold $Q = 0$.

The S--wave amplitude Eq.(\ref{label1.3}) can be represented in the
following form
\begin{eqnarray}\label{label1.5}
f^{K^-p}_0(Q) &=& \frac{1}{2iQ}\,\Big(\eta^{K^-p}_0(Q)\,
e^{\textstyle\,2i\delta^{K^-p}_0(Q)} - 1\Big) =\nonumber\\ &=&
\frac{1}{2iQ}\,\Big( e^{\textstyle\,2i\delta^{K^-p}_B(Q)} - 1\Big) +
e^{\textstyle\,2i\delta^{K^-p}_B(Q)}f^{K^-p}_0(Q)_R,
\end{eqnarray}
where $\delta^{K^-p}_B(Q)$ is the phase shift of an elastic 
background of low--energy $K^-p$ scattering and $f^{K^-p}_0(Q)_R$ is
the contribution of resonances. 

In our model of strong low--energy $\bar{K}N$ interactions near
threshold proposed in Ref.\cite{IV3} the imaginary part ${\cal
  I}m\,f^{K^0}_0(0)$ of the S--wave amplitude of $K^-p$ scattering is
defined by the contributions of strange baryon resonances
$\Lambda(1405)$, $\Lambda(1800)$ and $\Sigma(1750)$. This implies that
\begin{eqnarray}\label{label1.6}
{\cal I}m\,f^{K^-p}_0(0) = {\cal I}m\,f^{K^-p}_0(0)_R.
\end{eqnarray}
According to Gell--Mann's $SU(3)$ classification of hadrons, the
$\Lambda(1405)$ resonance is an $SU(3)$ singlet, whereas the
resonances $\Lambda(1800)$ and $\Sigma(1750)$ are components of an
$SU(3)$ octet Ref.\cite{PDG04}.

The real part ${\cal R}e\,f^{K^-p}_0(0)$ of the S--wave
amplitude of $K^-p$ scattering at threshold
\begin{eqnarray}\label{label1.7}
{\cal R}e\,f^{K^-p}_0(0) = {\cal R}e\,f^{K^-p}_0(0)_R
+ {\cal R}e\,\tilde{f}^{K^-p}_0(0)
\end{eqnarray}
is defined by the contribution of (i) the strange baryon resonances
${\cal R}e\,f^{K^-p}_0(0)_R$ in the s--channel of low--energy elastic
$K^-p$ scattering, (ii) the exotic four--quark (or $K\bar{K}$
molecules) scalar states $a_0(980)$ and $f_0(980)$ in the $t$--channel
of low--energy elastic $K^- p$ scattering and (iii) hadrons with
non--exotic quark structures, i.e.  $q\bar{q}$ for mesons and $qqq$
for baryons, where $q = u, d$ or $s$ quarks. The
contributions of exotic mesons and non--exotic hadrons we denote as
${\cal R}e\,\tilde{f}^{K^-p}_0(0)$.

According to Ref.\cite{MR96}, we describe strange baryon resonances as
elementary particle fields coupled to octets of low-lying baryons $B =
(N,\Lambda^0,\Sigma, \Xi)$ and pseudoscalar mesons $P =
(\pi,K,\bar{K},\eta(550))$. The effective phenomenological low--energy
Lagrangians of these interactions are Ref.\cite{IV3}:
\begin{eqnarray}\label{label1.8}
\hspace{-0.3in}{\cal L}_{\Lambda_1BP}(x) &=& g_1\bar{\Lambda}^0_1(x)\,{\rm
tr}\{B(x)P(x)\} + {\rm h.c.}, \nonumber\\ \hspace{-0.3in}{\cal L}_{B_2BP}(x) &=&
\frac{1}{\sqrt{2}}\,g_2\,{\rm tr}\{\{\bar{B}_2(x),B(x)\}P(x)\} +
\frac{1}{\sqrt{2}}\,f_2\,{\rm tr}\{[\bar{B}_2(x),B(x)]P(x)\} + {\rm h.c.},
\end{eqnarray}
where $g_1$, $g_2$ and $f_2$ are phenomenological coupling constants,
$\Lambda^0_1(x)$ and $B_2(x)$ are interpolating field operators of the
singlet $\Lambda(1405)$ and octet of strange baryon resonances,
respectively. The interactions of resonances with the meson--baryon
pairs $\bar{K}N$, $Y\pi$ and $Y\eta(550)$, where $Y = \Sigma^{\pm},
\Sigma^0$ or $\Lambda^0$, are given by
\begin{eqnarray}\label{label1.9}
{\cal L}_{\Lambda^0_1BP}(x) &=&
g_1\,\bar{\Lambda}^0_1(x)(\vec{\Sigma}(x)\cdot \vec{\pi}(x) -
p(x)K^-(x) + n(x)\bar{K}^0(x) + \frac{1}{3}\,\Lambda^0(x)\eta(x)) +
{\rm h.c.},\nonumber\\ {\cal L}_{\Lambda^0_2BP}(x) &=&
\frac{g_2}{\sqrt{3}}\,\bar{\Lambda}^0_2(x)(\vec{\Sigma}(x)\cdot
\vec{\pi}(x) - \Lambda^0(x)\eta(x))\nonumber\\ &+& \frac{g_2 +
3f_2}{2\sqrt{3}}\,\bar{\Lambda}^0_2(x)\,(p(x)K^-(x) -
n(x)\bar{K}^0(x)) + {\rm h.c.},\nonumber\\ {\cal
L}_{\Sigma^0_2BP}(x)&=& f_2\,\bar{\Sigma}^0_2(x)\,(\Sigma^-(x)\pi^+(x)
- \Sigma^+(x)\pi^-(x))\nonumber\\ &+&
\frac{g_2}{\sqrt{3}}\,\bar{\Sigma}^0_2(x)\,(\Lambda^0(x)\pi^0(x) +
\Sigma^0(x)\eta(x))\nonumber\\ &+& \frac{g_2 -
f_2}{2}\,\bar{\Sigma}^0_2(x)\,(- p(x)K^-(x) - n(x)\bar{K}^0(x)) + {\rm
h.c.},\nonumber\\
{\cal L}_{\Sigma^-_2BP}(x) &=& f_2\,\bar{\Sigma}^-_2(x) (\Sigma^-(x)
\pi^0(x) - \Sigma^0(x) \pi^-(x)) +
\frac{g_2}{\sqrt{3}}\,\bar{\Sigma}^-_2(x)\Lambda^0(x)\pi^-(x)\nonumber\\
&&- \frac{1}{\sqrt{2}}\,(g_2 - f_2)\,\bar{\Sigma}^-_2(x) n(x) K^-(x) +
{\rm h.c.}
\end{eqnarray}
As has been pointed out in Ref.\cite{MR96}, the inclusion of the
$\Lambda(1405)$ resonance as an elementary particle fields does not
contradict ChPT by Gasser and Leutwyler Ref.\cite{JG83} and allows to
calculate the low--energy parameters of $\bar{K}N$ scattering to
leading order in Effective Chiral Lagrangians.

Using the effective Lagrangians Eq.(\ref{label1.9}) we obtain the
S--wave amplitudes of inelastic channels of $K^-p$ scattering at
threshold $f(K^-p \to Y\pi)$, where $Y\pi = \Sigma^{\mp}\pi^{\pm}$,
$\Sigma^0\pi^0$ and $\Lambda^0\pi^0$. The theoretical cross sections
$\sigma(K^-p \to Y\pi)$ for these reactions satisfy the experimental
data Ref.\cite{DT71} 
\begin{eqnarray}\label{label1.10}
\gamma &=& \frac{\sigma(K^-p \to \Sigma^-\pi^+)}{\sigma(K^-p \to
\Sigma^+\pi^-)} = 2.360 \pm 0.040,\nonumber\\ R_c &=&
\frac{\sigma(K^-p \to \Sigma^-\pi^+) + \sigma(K^-p \to
\Sigma^+\pi^-)}{\sigma(K^-p \to \Sigma^-\pi^+) + \sigma(K^-p \to
\Sigma^+\pi^-) + \sigma(K^-p \to \Sigma^0\pi^0) + \sigma(K^-p \to
\Lambda^0\pi^0)} = \nonumber\\ &=&0.664 \pm 0.011,\nonumber\\ R_n &=&
\frac{\sigma(K^-p \to \Lambda^0\pi^0)}{\sigma(K^-p \to \Sigma^0\pi^0)
+ \sigma(K^-p \to \Lambda^0\pi^0)} = 0.189 \pm 0.015
\end{eqnarray}
with an accuracy of about $6\,\%$ and the constraint that the
$\Lambda(1800)$ resonance decouples from the $K^-p$ pair that gives
$f_2 = -g_2/3$.  This result is obtained without the specification of
the numerical values of the coupling constants $g_1$ and $g_2$ and the
masses of resonances, but using only physical masses of interacting
particles for the calculation of phase volumes. For $f_2 = -g_2/3$ the
S--wave amplitudes $f(K^-p\to Y\pi)$ can be defined by
\begin{eqnarray}\label{label1.11}
  f(K^-p \to \Sigma^-\pi^+) &=&+\,
  \frac{1}{4\pi}\,\frac{\mu}{~m_{K^-}}\,\sqrt{\frac{m_{\Sigma^-}}{m_p}}
  \Big(-\,A + \frac{1}{2}\,B\Big),\nonumber\\ f(K^-p \to
  \Sigma^+\pi^-) &=&+\, \frac{1}{4\pi}\,\frac{\mu}{~m_{K^-}}\,
  \sqrt{\frac{m_{\Sigma^+}}{m_p}}\,\Big(-\,A - \frac{1}{2}\,B\Big),\nonumber\\
  f(K^-p \to \Sigma^0\pi^0) &=&-\,
  \frac{1}{4\pi}\,\frac{\mu}{~m_{K^-}}\,
  \sqrt{\frac{m_{\Sigma^0}}{m_p}}\,A,\nonumber\\ 
  f(K^-p \to \Lambda^0\pi^0) &=&
  -\,\frac{1}{4\pi}\,
  \frac{\mu}{~m_{K^-}}\,\sqrt{\frac{m_{\Lambda^0}}{m_p}}\,\frac{\sqrt{3}}{2}\,B,
\end{eqnarray}
where $A = -\,6.02\,{\rm fm}$ is the contribution of the
$\Lambda(1405)$ resonance, calculated for $g_1 = 0.91$ and
$m_{\Lambda(1405)} = 1405\,{\rm MeV}$ Ref.\cite{IV3}.

The parameter $B$ describes the contribution of the baryon resonance
octet.  Unfortunately, in Ref.\cite{IV3} we have exaggerated the role
of the $\Sigma(1750)$ resonance in strong low--energy $\bar{K}N$
interactions at threshold. The contribution of the $\Sigma(1750)$
resonance with the recommended values of its parameters does not
define $B$ correctly. More definitely the contribution of the
$\Sigma(1750)$ resonance does not saturate the sum rule
Eq.(\ref{label2.13}), which is the consequence of the low--energy
theorem $a^0_0 + 3\,a^1_0 = 0$ Eq.(\ref{label2.8}). 

Therefore, instead of the assertion that $B$ is caused by the
contribution of the $\Sigma(1750)$ resonance we argue that {\it $B$ is
  defined by a contribution of a baryon  background with a
  property of an $SU(3)$ octet and quantum numbers of the
  $\Lambda(1800)$ and $\Sigma(1750)$ resonances}.  The former is
important for the correct description of the experimental data
Eq.(\ref{label1.10}).

Using the relation between the S--wave amplitudes of the reactions
$K^-p \to \Sigma^-\pi^+$ and $K^-p \to \Sigma^+\pi^-$, imposed by the
experimental data Eq.(\ref{label1.10}), we obtain the contribution of
the baryon background $B$ in terms of $\gamma$, $A$ and the phase
volumes of the final $\Sigma^-\pi^+$ and $\Sigma^+\pi^-$ states. This
gives
\begin{eqnarray}\label{label1.12}
  B = 2\,\frac{\sqrt{\gamma\,k_{\Sigma^+\pi^-}} - \sqrt{k_{\Sigma^-\pi^+}}}{
\sqrt{\gamma\,k_{\Sigma^+\pi^-}} + \sqrt{k_{\Sigma^-\pi^+}}}\,(-\,A) =
 2.68\,{\rm fm},
\end{eqnarray}
where $k_{\Sigma^+\pi^-} = 181.34\,{\rm MeV}$ and $k_{\Sigma^-\pi^+} =
172.73\,{\rm MeV}$ are the relative momenta of the
$\Sigma^{\pm}\pi^{\mp}$ pairs at threshold of $K^-p$ scattering,
calculated for physical masses of interacting particles
Ref.\cite{PDG04}. The phase volumes of the final $\Sigma^-\pi^+$ and
$\Sigma^+\pi^-$ states are equal to $k_{\Sigma^-\pi^+} /4\pi(m_K +
m_p)$ and $k_{\Sigma^+\pi^-}/4\pi(m_K + m_p)$, respectively.

The paper is organised as follows. In Section 2 we calculate the
S--wave amplitudes of $K^-p$ and $K^-n$ scattering at threshold. We
show that the S--wave scattering lengths $a^{K^-p}_0$ and $a^{K^-n}_0$
of $K^-p$ and $K^-n$ scattering satisfy the low--energy theorem
$a^{K^-p}_0 + a^{K^-n} = (a^0_0 + 3\,a^1_0)/2 = 0$. We show that in
the chiral limit due to isospin invariance $a^{K^-p}_0 + a^{K^-n} =
(a^0_0 + 3\,a^1_0)/2 = -\,\sqrt{6}\,b^0_0 = 0$, where $b^0_0 =
(a^{\pi^-p}_0 + a^{\pi^-n}_0)/2$ is the isoscalar S--wave scattering
length $\pi^-N$ scattering, vanishing in the chiral limit. The
low--energy theorem $a^0_0 + 3\,a^1_0 = 0$ can be also derived using
invariance of strong low--energy interactions under $U$--spin
rotations \cite{HL65}. We calculate the energy level displacement of
the ground state of kaonic hydrogen.  Theoretical value agrees well
with the experimental data by the DEAR Collaboration.  Using the
results obtained in Section 2 for the S--wave scattering lengths of
$K^-N$ scattering and in Ref.\cite{IV4} we recalculate the S--wave
scattering length $a^{K^-d}_0$ of $K^-d$ scattering at threshold.  We
calculate the energy level displacement of the ground state of kaonic
deuterium.  All results agree well with those obtained in
Ref.\cite{IV4}.  In Section 3 we analyse the cross sections for
elastic and inelastic $K^-p$ scattering for laboratory momenta
$70\,{\rm MeV}/c \le 150\,{\rm MeV}/c$ of the incident
$K^-$--meson. The theoretical cross sections agree with the available
experimental data within two standard deviations.  In the Conclusion
we discuss the obtained results.

\section{S--wave amplitude of $K^-N$ scattering at threshold}
\setcounter{equation}{0}

\subsection{S--wave amplitude of $K^-p$ scattering at threshold}

As has been shown in Ref.\cite{IV3}, the imaginary part of the S--wave
amplitude of $K^-p$ scattering at threshold can be represented by
\begin{eqnarray}\label{label2.1}
{\cal I}m\,f^{K^-p}_0(0) &=& {\cal I}m\,f^{K^-p}_0(0)_R = \frac{1}{R_c}\,
\Big(1 +
\frac{1}{\gamma}\Big)\,|f(K^-p \to \Sigma^-\pi^+)|^2
k_{\Sigma^-\pi^+} =\nonumber\\
&=& (0.35\pm 0.02)\,{\rm fm},
\end{eqnarray}
where $f(K^-p \to \Sigma^-\pi^+) = 0.43\,{\rm fm}$ and $\pm\,0.02$ is
an accuracy of about $6\,\%$ of our description of the experimental
data Eq.(\ref{label1.11}). The contribution of the $\Lambda(1405)$
resonance and the baryon background to ${\cal R}e\,f^{K^-0}_0(0)$ is
equal to Ref.\cite{IV3}
\begin{eqnarray}\label{label2.2}
{\cal R}e\,f^{K^-p}_0(0)_R = 
\frac{1}{4\pi}\,\frac{\mu}{~~m_K}\,(A + B) = (-\,0.17\pm 0.01)\,{\rm fm}.
\end{eqnarray}
Since the contribution ${\cal R}e\,\tilde{f}^{K^-p}_0(0) = (-\,0.33\pm
0.04)\,{\rm fm}$, calculated in Ref.\cite{IV3}, is not changed, the
total real part of the S--wave amplitude of $K^-p$ scattering at
threshold amounts to
\begin{eqnarray}\label{label2.3}
{\cal R}e\,f^{K^-p}_0(0) = {\cal R}e\,f^{K^-p}_0(0)_R
+{\cal R}e\,\tilde{f}^{K^-p}_0(0) = (-\,0.50 \pm 0.05)\,{\rm fm}.
\end{eqnarray}
Hence, for the S--wave amplitude of $K^-p$ scattering at threshold we get 
\begin{eqnarray}\label{label2.4}
f^{K^-p}_0(0) =  (-\,0.50\pm 0.05) + \,i\,(0.35 \pm 0.02)\,{\rm fm}.
\end{eqnarray}
This agrees well with the result obtained in Ref.\cite{IV3}.

\subsection{S--wave amplitude of $K^-n$ scattering at threshold}

Since the $K^-n$ pair has isospin $I = 1$, in our model the resonant
parts of the S--wave amplitudes of elastic and inelastic $K^-n$
scattering at threshold are described by the contribution of the
baryon background $B$.  The imaginary part ${\cal I}m\,f^{K^-n}_0(0)$
is defined by the inelastic channels $K^-n \to Y\pi$ with $Y\pi =
\Sigma^-\pi^0$, $\Sigma^0\pi^-$ and $\Lambda^0\pi^-$. Using the
results obtained in Ref.\cite{IV4} we get
\begin{eqnarray}\label{label2.5}
  {\cal R}e\,f^{K^-n}_0(0) &=& {\cal R}e\,\tilde{f}^{K^-n}_0(0) +  
  \frac{1}{2\pi}\,
  \frac{\mu}{m_K}\,B =\nonumber\\
  &=& (0.22\pm 0.02) + \frac{1}{2\pi}\,
  \frac{\mu}{m_K}\,B = (0.50 \pm 0.02)\,{\rm fm},\nonumber\\
  {\cal I}m\,f^{K^-n}_0(0) &=& \sum_{Y\pi}|f(K^-n \to Y\pi)|^2 
  k_{Y\pi} = 0.04
  \,{\rm fm},\nonumber\\
  f(K^-n \to \Sigma^-\pi^0)&=& +\,\frac{1}{4\pi}\, 
  \frac{\mu}{m_K}\,\sqrt{\frac{m_{\Sigma^-}}{m_p}}\,
  \frac{1}{\sqrt{2}}\,B = +\,0.11\,{\rm fm},
  \nonumber\\
  f(K^-n \to \Sigma^0\pi^-)&=& -\,\frac{1}{4\pi}\,
  \frac{\mu}{m_K}\,\sqrt{\frac{m_{\Sigma^0}}{m_p}}\,\frac{1}{\sqrt{2}}\,
  B = -\,0.11\,{\rm fm},\nonumber\\
  f(K^-n \to \Lambda^0\pi^-)&=& -\,\frac{1}{4\pi}\,
  \frac{\mu}{m_K}\,\sqrt{\frac{m_{\Lambda^0}}{m_p}}\,\frac{\sqrt{3}}{2}\,
  B = -\,0.13\,{\rm fm},
\end{eqnarray}
where $k_{\Sigma^-\pi^0} = 181.36\,{\rm MeV}$, $k_{\Sigma^0\pi^-} =
183.50\,{\rm MeV}$ and $k_{\Lambda^0\pi^-} = 256.88\,{\rm MeV}$ are
the relative momenta of the pairs $\Sigma^-\pi^0$, $\Sigma^0\pi^-$ and
$\Lambda^0\pi^-$ at threshold of $K^-n$ scattering. Since the
contribution of the exotic scalar mesons $a_0(980)$ and $f_0(980)$ to
the S--wave scattering amplitude of $K^-n$ scattering at threshold
vanishes, ${\cal R}e\,\tilde{f}^{K^-n}_0(0) = (0.22\pm 0.02)\,{\rm
  fm}$ is defined by low--energy interactions of non--exotic hadrons
only \cite{IV4}.

The S--wave amplitude of $K^-n$ scattering at threshold is equal to
\begin{eqnarray}\label{label2.6}
f^{K^-n}_0(0) =  (+\,0.50\pm 0.02)  + \,i\,(0.04 \pm 0.00)\,{\rm fm}.
\end{eqnarray}
Equating $f^{K^-p}_0(0) = (\tilde{a}^0_0 + \tilde{a}^1_0)/2$ and
$f^{K^-n}_0(0) = \tilde{a}^1_0$, where $\tilde{a}^0_0$ and
$\tilde{a}^1_0$ are complex S--wave scattering lengths of $\bar{K}N$
scattering with isospin $I = 0$ and $I = 1$, we get the numerical
values of $\tilde{a}^0_0$ and $\tilde{a}^1_0$:
\begin{eqnarray}\label{label2.7}
  \tilde{a}^0_0 &=& 
  (-\,1.50\pm 0.05)  + \,i(\,0.66 \pm 0.04)\,{\rm fm},
  \nonumber\\
  \tilde{a}^1_0  &=&
  (+\,0.50\pm 0.02)   + \,i\,(0.04 \pm 0.00)\,{\rm fm},
\end{eqnarray}
where ${\cal R}e\,\tilde{a}^0_0 = a^0_0 = (-\,1.50\pm 0.05)\,{\rm
  fm}$ and ${\cal R}e\,\tilde{a}^1_0 = a^1_0 = (+\,0.50\pm
0.02)\,{\rm fm}$.  

The complex S--wave scattering length $\tilde{a}^0_0$ agrees well with
the scattering length obtained by Dalitz and Deloff Ref.\cite{RD91}
 
\[\tilde{a}^0_0 = (-\,1.54 \pm 0.05) + \,i\,(0.74 \pm 0.02)\,{\rm fm}\]

\noindent for the position of the pole on sheet II of the $E$--plane
$E^* - i\,\Gamma/2$ with $E^* = 1404.9\,{\rm MeV}$ and $\Gamma =
53.1\,{\rm MeV}$ Ref.\cite{RD91}. This corresponds to our choice of
the parameters of the $\Lambda(1405)$ resonance. 

The complex S--wave scattering lengths Eq.(\ref{label2.7}) we apply to
the calculation of the energy level displacement of the ground state
of kaonic hydrogen. The real parts of these scattering lengths $a^0_0$
and $a^1_0$ we use for the calculation of the energy level shift of
the ground state of kaonic deuterium.

\subsection{Low--energy theorem  $a^0_0 + 3 a^1_0 = 0$}

The numerical values of the real parts of the S--wave scattering
lengths $a^{K^-p}_0= (a^0_0 + a^1_0)/2$ and $a^{K^-n}_0 = a^1_0$ of
$K^-N$ scattering satisfy the relation
\begin{eqnarray}\label{label2.8}
 a^{K^-p}_0 + a^{K^-n}_0 = \frac{1}{2}\,(a^0_0 + 3\,a^1_0) = 0.
\end{eqnarray}
This is the low--energy theorem valid in the chiral limit, which can
be derived relating the S--wave scattering lengths of $K^-N$
scattering to the S--wave scattering lengths of $\pi^- N$ scattering.

As has been shown by Weinberg Ref.\cite{SW66}, in the chiral limit the
S--wave scattering lengths $a^{\pi^-p}_0 = (2\,a^{1/2}_0 +
a^{3/2}_0)/3$ and $a^{\pi^-n}_0 = a^{3/2}_0$ of $\pi^-N$ elastic
scattering, where $a^{1/2}_0$ and $a^{3/2}_0$ are the S--wave
scattering lengths of $\pi N$ scattering with isospin $I = 1/2$ and $I
= 3/2$, obey the constraint
\begin{eqnarray}\label{label2.9}
  a^{\pi^-p}_0 + a^{\pi^-n}_0 = 
  \frac{2}{3}\,(a^{1/2}_0 + 2a^{3/2}) = 2\,b^0_0 = 0,
\end{eqnarray}
which is caused by Adler's consistency condition Ref.\cite{SA65},
where $b^0_0$ is the S--wave scattering length of $\pi N$ scattering
in the $t$--channel with isospin $I = 0$.

For the derivation of the low--energy theorem Eq.(\ref{label2.8}) it
is convenient to use the $\mathbb{K}$--matrix approach
Refs.\cite{TE88,VV04}.  In terms of the matrix elements of the
$\mathbb{K}$--matrix in the $t$--channel the sum of the S--wave
scattering lengths $a^{\pi^-p}_0 + a^{\pi^-n}_0$ is equal to
\begin{eqnarray}\label{label2.10}
 a^{\pi^-p}_0 + a^{\pi^-n}_0  = \langle \pi^+\pi^-|\mathbb{K}|\bar{p}p +
 \bar{n}n\rangle = 
  -\,\sqrt{\frac{2}{3}}\,\langle I = 0|\mathbb{K}|I = 0\rangle,
\end{eqnarray}
where we have taken into account the isospin properties of the
hadronic state $|\bar{p}p + \bar{n}n\rangle$ and $|\pi^-\pi^+\rangle$.
Setting
\begin{eqnarray}\label{label2.11}
  \langle I = 0|\mathbb{K}|I = 0\rangle = -\,\sqrt{6}\,b^0_0
\end{eqnarray}
we arrive at the low--energy theorem Eq.(\ref{label2.9}).

In terms of the matrix element of the $\mathbb{K}$--matrix in the
$t$--channel the sum of the S--wave scattering lengths $a^{K^-p}_0 +
a^{K^-n}_0$ can be defined by
\begin{eqnarray}\label{label2.12}
  a^{K^-p}_0 + a^{K^-n}_0  =  \langle K^+K^-|\mathbb{K}|\bar{p}p + 
  \bar{n}n\rangle = \langle I = 0|\mathbb{K}|I = 0\rangle = 
-\,\sqrt{6}\,b^0_0 = 0.
\end{eqnarray}
This proves the low--energy theorem Eq.(\ref{label2.8}), which is, of
course, valid only at leading order in chiral expansion. The former
becomes more obvious if to derive the low--energy theorem
Eq.(\ref{label2.8}) using invariance of strong low--energy
interactions under $U$--spin rotations Ref.\cite{HL65}. According to
$U$--spin classification of the components of the pseudoscalar octet
\cite{HL65}, the mesons $\pi$ and $K$ transform as components of
doublets $(K^+,\pi^+)$ and $(\pi^-,K^-)$. This can be allowed only
in the chiral limit.

The relation Eq.(\ref{label2.8}) can be rewritten in the form of the
sum rule
\begin{eqnarray}\label{label2.13}
 {\cal R}e\tilde{f}^{K^-p}_0(0) +
  {\cal R}e\tilde{f}^{K^-n}_0(0) = -\,\frac{1}{4\pi}\,\frac{\mu}{m_{K^-}}\,
(A + 3 B).
\end{eqnarray}
Using the numerical values ${\cal R}e\tilde{f}^{K^-p}_0(0) =
-\,0.33\,{\rm fm}$, ${\cal R}e\tilde{f}^{K^-n}_0(0) = 0.22\,{\rm fm}$,
$A = -\,6.02\,{\rm fm}$ and $B = 2.68\,{\rm fm}$, one can show that the sum 
rule
Eq.(\ref{label2.13}) is fulfilled
\begin{eqnarray*}
  {\cal R}e\tilde{f}^{K^-p}_0(0) +
  {\cal R}e\tilde{f}^{K^-n}_0(0) = -\,0.11\,{\rm
    fm}\quad,\quad
  -\,\frac{1}{4\pi}\,\frac{\mu}{m_{K^-}}\,
  (A + 3 B) =  -\,0.11\,{\rm fm}.
\end{eqnarray*}
Unfortunately, the $\Sigma(1750)$ resonance does not saturate the sum
rule Eq.(\ref{label2.13}).

In our model of strong $\bar{K}N$ interactions at threshold the l.h.s.
of Eq.(\ref{label2.13}) is defined by quark--hadron interactions,
whereas the r.h.s. of Eq.(\ref{label2.13}) is the resonant part,
caused by the contribution of the $\Lambda(1405)$ resonance $A$ and
the baryon background $B$.  This is to some extent a manifestation of
quark--hadron duality pointed out by Shifman {\it et al.} within
non--perturbative QCD in the form QCD sum rules \cite{SVZ}. Since the
l.h.s. of Eq.(\ref{label2.13}) can be calculated independently of the
assumption of the contribution of the $\Lambda(1405)$ resonance and
the baryon background, the sum rule Eq.(\ref{label2.13}) places
constraints on the parameters of the $\Lambda(1405)$ resonance and the
baryon background calculated at leading order in chiral expansion.

\subsection{Energy level displacement of the ground state of kaonic
  hydrogen}

For the S--wave amplitude of $K^-p$ scattering at threshold
Eq.(\ref{label2.4}) the energy level displacement of the ground state
of kaonic hydrogen is equal to
\begin{eqnarray}\label{label2.14}
  - \,\epsilon^{(0)}_{1s} + i\,\frac{\Gamma^{(0)}_{1s}}{2} = 
421.13\,f^{K^-p}_0(0) = 421.13\;\frac{\tilde{a}^0_0 + 
\tilde{a}^1_0}{2} = (-\,205 \pm 21) 
+ \,i\,(144 \pm 9)\,{\rm eV}.~~~ 
\end{eqnarray}
This result agrees well with the experimental data by the DEAR
Collaboration Eq.(\ref{label1.1}). 

As has been shown in Ref.\cite{IV6}, the energy level shift and width
of the ground state of kaonic hydrogen acquire the dispersive
corrections, caused by the intermediate $\bar{K}^0n$ state on--mass
shell
\begin{eqnarray}\label{label2.15}
  \hspace{-0.3in}\delta^{Disp}_S &=& \frac{\delta 
    \epsilon^{\bar{K}^0n}_{1s}}{\epsilon^{(0)}_{1s}} = \frac{1}{4}\,
  (a^1_0 - a^0_0)^2\,q^2_0
  = (8.6\pm 0.9)\,\%,\nonumber\\
  \hspace{-0.3in}\delta^{Disp}_W &=& \frac{\delta 
    \Gamma^{\bar{K}^0n}_{1s}}{\Gamma^{(0)}_{1s}} = 
  \frac{1}{2\pi}\,
  \frac{(a^1_0 -a^0_0)^2}{{\cal I}m\,f^{K^-p}_0(0)\,a_B}\,
  {\ell n}\Big[\frac{2 a_B}{|a^0_0 + a^1_0|}\Big] = (11.1\pm 1.2)\,\%,
\end{eqnarray}
where $q_0 = \sqrt{2\mu(m_{\bar{K}^0} - m_{K^-} + m_n - m_p)} =
58.35\,{\rm MeV}$, calculated for $m_{\bar{K}^0} - m_{K^-} =
3.97\,{\rm MeV}$ and $ m_n - m_p = 1.29\,{\rm MeV}$ Ref.\cite{PDG04} and
$a_B = 1/\alpha \mu = 83.59\,{\rm fm}$ is the Bohr radius.

Taking into account the dispersive corrections
Eq.(\ref{label2.15}), the energy level displacement of the ground state
of kaonic hydrogen is equal to
\begin{eqnarray}\label{label2.16}
  - \,\epsilon^{(\rm th)}_{1s} + i\,\frac{\Gamma^{(\rm th)}_{1s}}{2} = 
  (- 223 \pm 21) + \,i\,(159 \pm 9)\,{\rm eV}. 
\end{eqnarray}
As we have shown above that the S--wave scattering lengths of $K^-p$
and $K^-n$ scattering are calculated at leading order in chiral
expansion and satisfy the low--energy theorem Eq.(\ref{label2.8}).
This allows to take into account contributions, caused by
next--to--leading order corrections in chiral expansion.

The most important next--to--leading order correction in chiral
expansion is the contribution of the $\sigma^{(I = 1)}_{KN}(0)$--term,
given by Ref.\cite{IV5}:
\begin{eqnarray}\label{label2.17}
   \delta \epsilon^{(\sigma)}_{1s} =  \frac{\alpha^3 \mu^3}{2\pi m_K 
F^2_K}\Big[\sigma^{(I = 1)}_{KN}(0) - 
  \frac{m^2_K}{4m_N}i\int d^4x\langle p(\vec{0},\sigma_p)
  |{\rm T}(J^{4+i5}_{50}(x)J^{4-i5}_{50}(0))|
  p(\vec{0},\sigma_p)\rangle\Big].\quad
\end{eqnarray} 
Here $J^{4\pm i5}_{50}(x)$ are time--components of the axial--vector
hadronic currents $J^{4\pm i5}_{5\mu}(x)$, changing strangeness
$|\Delta S| = 1$, $F_K = 113\,{\rm MeV}$ is the PCAC constant of the
$K$--meson Ref.\cite{PDG04} and the $\sigma^{(I = 1)}_{KN}(0)$--term is
defined by Refs.\cite{ER72}--\cite{JG99} 
\begin{eqnarray}\label{label2.18}
  \sigma^{(I = 1)}_{KN}(0) = \frac{m_u + m_s}{4m_N}\,
\langle p(\vec{0},\sigma_p)|\bar{u}(0)u(0) + \bar{s}(0)s(0)|p(\vec{0},
\sigma_p)\rangle,
\end{eqnarray}
where $u(0)$ and $s(0)$ are operators of the interpolating fields of
$u$ and $s$ current quarks Ref.\cite{FY83}.  

The correction $\delta \epsilon^{(\sigma)}_{1s}$ to the shift of the
energy level of the ground state of kaonic hydrogen, caused by the
$\sigma^{(I = 1)}_{KN}(0)$, is obtained from the S--wave amplitude of
$K^-p$ scattering, calculated to next--to--leading order in ChPT
expansion at the tree--hadron level Ref.\cite{IV5} and Current Algebra
Refs.\cite{SA68,HP73} (see also Ref.\cite{ER72}):
\begin{eqnarray}\label{label2.19}
 \hspace{-0.3in} 4\pi\,\Big(1 + \frac{m_K}{m_N}\Big)\,\tilde{f}^{K^-p}_0(0) &=& 
\frac{m_K}{F^2_K} - 
  \frac{1}{F^2_K}\,\sigma^{(I = 1)}_{KN}(0)\nonumber\\
 \hspace{-0.3in}  &+& 
  \frac{m^2_K}{4m_N F^2_K} i\int d^4x \langle p(\vec{0},\sigma_p)
  |{\rm T}(J^{4+i5}_{50}(x)J^{4-i5}_{50}(0))|
  p(\vec{0},\sigma_p)\rangle.\quad
\end{eqnarray} 
The contribution of the $\sigma^{(I = 1)}_{KN}(0)$--term, $-
\sigma^{(I = 1)}_{KN}(0)/F^2_K$, to the S--wave amplitude of $K^-p$
scattering in Eq.(\ref{label2.19}) has a standard structure
Ref.\cite{ER72}.

Since the first term $m_K/F^2_K,$ calculated to leading order in
chiral expansion, has been already taken into account in
Ref.\cite{IV3}, the second term, $- \sigma^{(I = 1)}_{KN}(0)/F^2_K$,
and the third one define next--to--leading order corrections in chiral
expansion to the S--wave amplitude of $K^-p$ scattering at threshold.

Taking into account the contribution $\delta
\epsilon^{(\sigma)}_{1s}$, the total shift of the energy level of the
ground state of kaonic hydrogen is equal to
\begin{eqnarray}\label{label2.20}
  \hspace{-0.3in} \epsilon^{(\rm th)}_{1s} &=& 223 \pm 21  +
  \frac{\alpha^3 \mu^3}{2\pi m_K 
    F^2_K} \sigma^{(I = 1)}_{KN}(0)\nonumber\\
\hspace{-0.3in}&-& \frac{\alpha^3 \mu^3 m_K}{ 
    8\pi  F^2_K m_N} i\!\int\! d^4x \langle p(\vec{0},\sigma_p)
  |{\rm T}(J^{4+i5}_{50}(x)J^{4-i5}_{50}(0))|
  p(\vec{0},\sigma_p)\rangle.
\end{eqnarray}
The theoretical estimates of the value of $\sigma^{(I = 1)}_{KN}(0)$,
carried out within ChPT with a dimensional regularization of divergent
integrals, are converged around the number $\sigma^{(I = 1)}_{KN}(0) =
(200 \pm 50)\,{\rm MeV}$ Refs.\cite{VB93,BB99}.  Hence, the contribution of
$\sigma^{(I = 1)}_{KN}(0)$ to the energy level shift amounts to
\begin{eqnarray}\label{label2.21}
 \frac{\alpha^3 \mu^3}{2\pi m_K 
    F^2_K}\,\sigma^{(I = 1)}_{KN}(0)  =  (67\pm 17)\,{\rm eV}.
\end{eqnarray}
The total shift of the energy level of the ground state of kaonic
hydrogen is given by
\begin{eqnarray}\label{label2.22}
   \epsilon^{(\rm th)}_{1s} =  (290  \pm 27) - 
  \frac{\alpha^3 \mu^3 m_K}{ 
    8\pi F^2_K m_N}\,i\int d^4x\,\langle p(\vec{0},\sigma_p)
  |{\rm T}(J^{4+i5}_{50}(x)J^{4-i5}_{50}(0))|
  p(\vec{0},\sigma_p)\rangle.
\end{eqnarray}
Hence the theoretical analysis of the second term in
Eq.(\ref{label2.22}) is required for the correct understanding of the
contribution of the $\sigma^{I = 1}_{KN}(0)$--term to the energy level shift.

Of course, one can solve the inverse problem. Indeed, calculating the
contribution of the term
\begin{eqnarray}\label{label2.23}
  \frac{\alpha^3 \mu^3 m_K}{ 
    8\pi F^2_K m_N}\,i\int d^4x\,\langle p(\vec{0},\sigma_p)
  |{\rm T}(J^{4+i5}_{50}(x)J^{4-i5}_{50}(0))|
  p(\vec{0},\sigma_p)\rangle
\end{eqnarray}
in Eq.(\ref{label2.20}) and using the experimental data on the energy
level shift, measured by the DEAR Collaboration Eq.(\ref{label1.1}),
one can extract the value of the $\sigma^{(I = 1)}_{KN}(0)$--term.

\subsection{Energy level displacement of the ground state of kaonic
  deuterium}

Using the real parts of the S--wave scattering lengths of $K^-N$
scattering Eq.(\ref{label2.7}) we recalculate the S--wave scattering
length $a^{K^-d}_0$ of $K^-d$ scattering. As has been
shown in Ref.\cite{IV4}, the S--wave scattering length $a^{K^-d}_0$ is
equal to
\begin{eqnarray}\label{label2.24}
  a^{K^-d}_0 = (a^{K^-d}_0)_{\rm EW} +  {\cal R}e\,\tilde{f}^{\,K^-d}_0(0),
\end{eqnarray}
where $(a^{K^-d}_0)_{\rm EW}$ is the Ericson--Weise scattering length
of $K^-d$ scattering in the S--wave state Ref.\cite{IV4}
\begin{eqnarray}\label{label2.25}
  (a^{K^-d}_0)_{\rm EW} &=&  \frac{1 + m_K/m_N}{1 + m_K/m_d}\,\frac{1}{2}\,(a^0_0 + 3 a^1_0) + \frac{1}{4}\,\Big(1 +
  \frac{m_K}{m_d}\Big)^{-1}\Big(1 + \frac{m_K}{m_N}\Big)^2\nonumber\\
  &&\times\,\Big(4a^1_0(a^0_0 + a^1_0) - (a^0_0 - a^1_0)^2\Big)\, \Big\langle \frac{1}{r_{12}}\Big\rangle,
\end{eqnarray}
where the term proportional to $(a^0_0 - a^1_0)^2$ describes the
contribution of the charge--exchange channel and $r_{12}$ is a
distance between two scatterers $n$ and $p$ Ref.\cite{TE88}. In our
approach $\langle 1/r_{12}\rangle$ is defined by Ref.\cite{IV4}
\begin{eqnarray}\label{label2.26}
\Big\langle \frac{1}{r_{12}}\Big\rangle = \int
d^3x\,\Psi^*_d(\vec{r}\,)\,\frac{\displaystyle e^{\textstyle\,- m_K
r}}{r}\,\Psi_d(\vec{r}\,) = 0.29\,m_{\pi},
\end{eqnarray}
where $\Psi_d(\vec{r}\,)$ is the wave function of the deuteron in the
ground state. 

We would like to remind that Ericson and Weise did not investigate the
$K^-d$ scattering. They analysed only $\pi^-d$
scattering \cite{TE88}. However, since the structure of the
contribution, given by Eq.(\ref{label2.25}), is very similar to that of
$\pi^-d$ scattering we call such a contribution as the Ericson--Weise
scattering length $(a^{K^-d}_0)_{\rm EW}$, which has been derived in
Ref.\cite{IV4} at the quantum field theoretic level.

The double scattering contribution to the S--wave amplitude of $K^-d$
scattering has been calculated by Kamalov {\it et al.} \cite{EO01}. In
the notation by Kamalov {\it et al.} the amplitude
$\tilde{f}^{\,K^-d}_0(0)_{\rm EW}$ reads
\begin{eqnarray}\label{label2.27}
  \tilde{f}^{\,K^-d}_0(0)_{\rm EW} = \Big(1 +
  \frac{m_K}{m_d}\Big)^{-1}\Big(1 + \frac{m_K}{m_N}\Big)^2\,(2 a_p a_n  - a^2_x)\, 
\Big\langle \frac{1}{r_{12}}\Big\rangle,
\end{eqnarray}
where $a_p = (a^0_0 + a^1_0)/2$, $a_n = a^1_0$ and $a_x = (a^1_0 -
a^0_0)/2$\,\footnote{In our former version nucl--th/0505078v1 the
  contribution of the double scattering contained the factor $(a_p a_n
  - a^2_x)$ instead of $(2 a_p a_n - a^2_x)$, where the term
  proportional to $a^2_x$ is defined by the charge--exchanged channel,
  \cite{IV4a}. We are grateful to Avraham Gal for calling our
  attention to this discrepancy. The replacement of $a_pa_n$ by $2a_p
  a_n$ changes the contribution of the double scattering and,
  correspondingly, the S--wave scattering length of $K^-d$ scattering
  by $17\,\%$, which is commensurable with a theoretical
  uncertainty.}.

The term ${\cal R}e\,\tilde{f}^{\,K^-d}_0(0)$ in Eq.(\ref{label2.24})
is defined by the inelastic two--body and three--body channels of the
$K^-d$ scattering at threshold. As has been shown in
Ref.\cite{IV4}, the contribution of this term is negligible in
comparison with the Ericson--Weise scattering length
$(a^{K^-d}_0)_{\rm EW}$. Dropping the contribution of this term we get
\begin{eqnarray}\label{label2.28}
  a^{K^-d}_0 = (a^{K^-d}_0)_{\rm EW} =  (-\,0.57\pm 0.07)\,{\rm fm}.
\end{eqnarray}
Since the imaginary part of the S--wave amplitude of $K^-d$ scattering
at threshold is not changed, using the results of Ref.\cite{IV4} we
obtain
\begin{eqnarray}\label{label2.29}
f^{K^-d}_0(0) = (-\,0.57 \pm 0.07) + i\,(0.52 \pm 0.08)\,{\rm fm}.
\end{eqnarray}
The energy level displacement for the ground state of
kaonic deuterium agrees well with that obtained in Ref.\cite{IV4}:
\begin{eqnarray}\label{label2.30}
-\,\epsilon_{1s} + i\,\frac{\Gamma_{1s}}{2} = 601.56\,f^{K^-d}_0(0) =
 (-\,343 \pm 42) + i\,(315 \pm 48)\;{\rm eV}.
\end{eqnarray}
The value of the S--wave amplitude of $K^-d$ scattering at threshold
as well as of the energy level displacement of the ground state of
kaonic deuterium agree well with the results obtained in
Ref.\cite{IV4}.

\section{Cross sections for low--energy $K^-p$ scattering}
\setcounter{equation}{0}

In this Section we apply our model of strong $K^-N$ interactions at
threshold to the description of the experimental data on the cross
sections for the reactions $K^-p \to K^-p$ and $K^-p \to Y\pi$, where
$Y\pi = \Sigma^{\mp}\pi^{\pm}$, $\Sigma^0\pi^0$ and $\Lambda^0\pi^0$,
as functions of a laboratory momentum $p_{lab}$ of the incident $K^-$
meson.  The available experimental data of the cross sections are
given for the laboratory momenta $50\,{\rm MeV}/c \le p_{lab} \le
250\,{\rm MeV}/c$ Ref.\cite{WW04}.  This corresponds to relative
momenta $40\,{\rm MeV}/c \le k \le 200\,{\rm MeV}/c$ of the $K^-p$
pair.

We analyse the cross sections for the reactions $K^-p \to K^-p$ and
$K^-p \to Y\pi$ only for the laboratory momenta $70\,{\rm MeV}/c \le
p_{lab} \le 150\,{\rm MeV}/c$ of the incident $K^-$, where the
experimental data are most reliable.  For these momenta the S--wave
amplitudes of the inelastic reactions $K^-p \to Y\pi$ are described
well by the S--wave scattering lengths
\begin{eqnarray}\label{label3.1}
  f(K^-p \to \Sigma^-\pi^+) &=& a_{\Sigma^-\pi^+} = +\,0.43\,{\rm fm}\,,\,
f(K^-p \to
  \Sigma^+\pi^-) = a_{\Sigma^+\pi^-} = +\,0.28\,{\rm fm},\nonumber\\
  f(K^-p \to \Sigma^0\pi^0) &=&a_{\Sigma^0\pi^0}= +\,0.36\,{\rm fm}\,,\,
  f(K^-p \to \Lambda^0\pi^0) = a_{\Lambda^0\pi^0}=-\,0.14\,{\rm fm}.
\end{eqnarray}
For laboratory momenta $70\,{\rm MeV}/c \le p_{lab} \le 150\,{\rm
  MeV}/c$, due to smallness of the S--wave scattering lengths
$a_{Y\pi}$, the cross sections are equal to
\begin{eqnarray}\label{label3.2}
  \sigma_{\Sigma^-\pi^+}(k) &=&4\pi\,\frac{k_{\Sigma^-\pi^+}(k)}{k}\,
C^2_0(k)\,a^2_{\Sigma^-\pi^+}\,,\,
\sigma_{\Sigma^+\pi^-}(k) = 4\pi\,\frac{k_{\Sigma^+\pi^-}(k)}{k}\,
C^2_0(k)\,a^2_{\Sigma^+\pi^-},
\nonumber\\
\sigma_{\Sigma^0\pi^0}(k) &=&4\pi\,\frac{k_{\Sigma^0\pi^0}(k)}{k}\,
C^2_0(k)\,a^2_{\Sigma^0\pi^0}\,,\,
 \sigma_{\Lambda^0\pi^0}(k) = 4\pi\,\frac{k_{\Lambda^0\pi^0}(k)}{k}\,
C^2_0(k)\,a^2_{\Lambda^0\pi^0},
\end{eqnarray}
where $C^2_0(k)$ is the contribution of the Coulomb interaction of the
$K^-p$ pair
\begin{eqnarray}\label{label3.3}
C^2_0(k) = \frac{2\pi \alpha \mu}{k}\,\frac{1}{\displaystyle 1 -
 e^{\textstyle -\,2\pi \alpha \mu/k}}.
\end{eqnarray}
The cross sections for inelastic channels agree well with those
obtained in Refs.\cite{RD60,RD62} (see also Ref.\cite{WH62})). The
calculation of the cross sections Eq.(\ref{label3.2}), taking into
account the Coulomb interaction in the initial and final state, one
can carry out within the potential model approach with strong
low--energy interactions described by the effective zero--range
potential Ref.\cite{IV7}:
\begin{eqnarray}\label{label3.4}
  V(\vec{r}\,) = - \frac{2\pi}{\mu}\,a_{Y\pi}\,\delta^{(3)}(\vec{r}\,),
\end{eqnarray}
where $a_{Y\pi}$ is a S--wave scattering length of the inelastic
channel under consideration. The S--wave amplitude
$f(\vec{k},\vec{k}_{Y\pi})$ of the inelastic channel $K^-p \to Y\pi$
is defined by the spatial integral
\begin{eqnarray}\label{label3.5}
  \hspace{-0.3in}f(\vec{k},\vec{k}_{Y\pi}) &=& -\,\frac{\mu}{2\pi}\int d^3x\,
  e^{\textstyle -\,i\,\vec{k}_{Y\pi}\cdot \vec{r}} V(\vec{r}\,)
  \psi^{\,C}_{K^-p}(\vec{k},\vec{r}\,)=\nonumber\\
  \hspace{-0.3in}&=& a_{Y\pi}\,e^{\textstyle\,\pi/2ka_B}\,\Gamma(1 - i/ka_B).
\end{eqnarray} 
Here $\psi^{\,C}_{K^-p}(\vec{k},\vec{r}\,)$ is the exact
non--relativistic Coulomb wave function of the relative motion of the
$K^-p$ pair in the incoming scattering state with a relative momentum
$\vec{k}$. It is given by Ref.\cite{LL65}
\begin{eqnarray}\label{label3.6}
  \psi^C_{K^-p}(\vec{k},\vec{r}\,) =  e^{\textstyle\,\pi/2ka_B}\,
  \Gamma(1 - i/ka_B)\,
  e^{\textstyle\,i\,\vec{k}\cdot \vec{r}}\,F(i/ka_B,1, ikr -
  i\,\vec{k}\cdot \vec{r}\,),
\end{eqnarray} 
where $F(i/ka_B,1, ikr - i\,\vec{k}\cdot \vec{r}\,)$ is the confluent
hypergeometric function Refs.\cite{LL65,MA72}.

The numerical values of the theoretical cross sections for the
reactions $K^-p \to \Sigma^-\pi^+$, $K^-p \to \Sigma^+\pi^-$, $K^-p
\to \Sigma^0\pi^0$ and $K^-p \to \Lambda^0\pi^0$, calculated for the
experimental values of the masses of interacting hadrons \cite{PDG04},
are adduced in Table 1 and the experimental data are given in Table 2
Ref.\cite{JC83} and Table 3 Ref.\cite{KT92}.  The cross sections as
functions of the laboratory momentum of the incident $K^-$ meson are
represented in Fig.1. It is seen that theoretical cross sections agree
with experimental data within two standard deviations.

For the S--wave scattering length $a_{\Lambda^0\pi^0} = -\,0.14\,{\rm
  fm}$ of the inelastic $K^-p \to \Lambda^0\pi^0$ reaction we
calculate the S--wave phase shift of $\Lambda\pi$ scattering at
threshold of the $\bar{K}N$ pair $\delta^{\Lambda^0\pi^0}_S =
a_{\Lambda^0\pi^0}k_{\Lambda^0\pi^0} = -\,10.3^0$. This agrees well
with recent results obtained by Tandean {\it et al.} Ref.\cite{JT01}
(see Fig. 3 and take the value of the phase shift of $\Lambda\pi$
scattering at threshold of the $\bar{K}N$ pair production).

Due to a contribution of the pure Coulomb interaction to the S--wave
amplitude of elastic $K^-p$ scattering only a differential cross
section for elastic $K^-p$ scattering is well defined.  For the
analysis of experimental data the differential cross section for
elastic $K^-p$ scattering has been taken in the form Ref.\cite{WH62}
(see also Refs.\cite{RD60,RD62}):
\begin{eqnarray}\label{label3.7}
\frac{d\sigma^{\,e\ell}_{pK^-}(k)}{d\Omega} = 
\Big|\frac{\sec^2(\theta/2)}{2k^2 a_B}\,
\exp\Big[\frac{2i}{ka_B}\,\sin(\theta/2)\Big] +
 C^2_0(k)R\,\exp(i\alpha)\Big|^2,
\end{eqnarray}
where $R$ and $\alpha$ are the experimental fit parameters
Ref.\cite{WH62}.

For the momenta $100\,{\rm MeV}/c \le p_{lab} \le 175\,{\rm MeV}/c$
the experimental values of the fit parameters, obtained in
Ref.\cite{WH62}, are $R = (0.81 \pm 0.06)\,{\rm fm}$ and $\alpha = (78
\pm 31)^0$. In our model the theoretical values of these parameters
are equal to $R = |a^{K^-p}_0| = (0.50 \pm 0.05)\,{\rm fm}$ and
$\alpha = 180^0$.  However, the experimental values for the cross
section for elastic $K^-p$ scattering, obtained in Ref.\cite{MS65} for
momenta $100\,{\rm MeV}/c \le p_{lab} \le 160\,{\rm MeV}/c$, by a
factor 1.5 smaller than the data by Humphrey and Ross Ref.\cite{WH62}.
This implies that the parameter $R$ can be reduced to the value $R
\approx 0.67$, which agrees better with our prediction.

The cross sections for elastic and inelastic $K^-p$ scattering
Eq.(\ref{label3.2}), defined for the momenta $70\,{\rm MeV/c} \le
p_{lab} \le 150\,{\rm MeV/c}$, do not contradict to the theoretical
results obtained by Borasoy {\it et al.} \cite{WW04}. The agreement of
the theoretical predictions for the cross sections of elastic and
inelastic $K^-p$ scattering is qualitative within about two standard
deviations. However, due to self--consistency of our calculation of
the S--wave amplitudes of $K^-N$ scattering at threshold and the
agreement with the experimental data by the DEAR Collaboration, we can
argue that the experimental values of the cross sections for elastic
and inelastic channels of $K^-p$ scattering as well as for $K^-n$
scattering should be remeasured. The same recommendation has been
pointed out by Borasoy {\it et al.}  Ref.\cite{WW04}.

\section{Conclusion}
\setcounter{equation}{0}

We have revisited our phenomenological quantum field theoretic model
of strong low--energy $\bar{K}N$ interactions at threshold. The main
change concerns the replacement of the contribution of the
$\Sigma(1750)$ resonance with quantum numbers $I\,(J^P) =
1\,(\frac{1}{2}^-)$ by the baryon background with the same quantum
numbers and $SU(3)$ properties. We remind that according to
Gell--Mann's classification of hadrons, the $\Sigma(1750)$ resonance
belongs to an $SU(3)$ octet of baryons.  Following our previous
analysis of strong low--energy $\bar{K}N$ interactions Ref.\cite{IV3}
and assuming that the S--wave amplitudes of inelastic channels of
$K^-p$ scattering at threshold are fully defined by the contribution
of the $\Lambda(1405)$ resonance with quantum numbers $I\,(J^P) =
0\,(\frac{1}{2}^-)$ and the octet of baryon background with $J^P =
\frac{1}{2}^-$ we describe the experimental data on ratios of the
cross sections for inelastic channels of $K^-p$ scattering
Eq.(\ref{label1.10}) within an accuracy of about $6\,\%$. Since the
non--resonant parts of the S--wave amplitudes are not changed, we have
used them and calculated the complex S--wave scattering lengths
$\tilde{a}^0_0$ and $\tilde{a}^1_0$ of $\bar{K}N$ scattering with
isospin $I = 0$ and $I = 1$, given by Eq.(\ref{label2.7}).
The complex S--wave scattering length $\tilde{a}^0_0$ agrees
well with that obtained by Dalitz and Deloff Ref.\cite{RD91}.

It is interesting to notice that the complex S--wave scattering length
$\tilde{a}^0_0$, calculated in our model, does not contradict the
result obtained by Akaishi and Yamazaki \cite{YA02} under the
assumption that the $\Lambda(1405)$ resonance is the bound $K^-p$
state.

The real parts of the complex S--wave scattering lengths $a^{K^-p}_0 =
(a^0_0 + a^1_0)/2$ and $a^{K^-n}_0 = a^1_0$ of $K^-N$ scattering
satisfy the low--energy theorem Eq.(\ref{label2.8}). As we have shown
above, this low--energy theorem is a $\bar{K}N$ scattering version of
the well--known low--energy theorem for the S--wave scattering lengths
of $\pi^-N$ scattering by Weinberg \cite{SW66}. 

The low--energy theorem Eq.(\ref{label2.8}) can be rewritten in the
form of the sum rule (\ref{label2.13}), where the l.h.s. is defined
by quark--hadron interactions, whereas the r.h.s. is the resonant part
caused by the contribution of the $\Lambda(1405)$ resonance $A$ and
the baryon background $B$. The sum rule Eq.(\ref{label2.13}) can be
accepted as a manifestation of quark--hadron duality pointed out by
Shifman {\it et al.} \cite{SVZ} within non--perturbative QCD in the
form of QCD sum rules.

Since in our model the S--wave scattering lengths are calculated to
leading order in chiral expansion and satisfy the low--energy theorem
Eq.(\ref{label2.8}), we can argue that our model is self--consistent
to leading order in chiral expansion. This implies that the inclusion
of the contributions of next--to--leading order corrections in chiral
expansion is well--defined and allows to provide the investigation of
the contribution of the $\sigma^{I = 1}_{KN}(0)$--term to the S--wave
scattering length of $K^-p$ and $K^-d$ scattering and the energy level
displacements of the ground states of kaonic atoms.

The analysis of the contribution of the $\sigma^{I = 1}_{KN}(0)$--term
demands the calculation of the quantity, defined by Eq.(\ref{label2.23}),
$$
\frac{\alpha^3 \mu^3 m_K}{
  8\pi F^2_K m_N}\,i\int d^4x\,\langle p(\vec{0},\sigma_p) |{\rm
  T}(J^{4+i5}_{50}(x)J^{4-i5}_{50}(0))|p(\vec{0},\sigma_p)\rangle.
$$
We are planning to carry out this calculation in our
forthcoming publication.

The energy level displacement of the ground state of kaonic hydrogen
Eq.(\ref{label2.14}), calculated for the complex S--wave scattering
lengths Eq.(\ref{label2.7}), agrees well with the result obtained in
Ref.\cite{IV3} and the experimental data by the DEAR
Collaboration. The account for the contribution of the dispersive
corrections, caused by the intermediate $\bar{K}^0n$ state on--mass
shell Ref.\cite{IV6}, changes the values of the energy level shift and
width by about $8\,\%$.

We have recalculated the S--wave scattering length $a^{K^-d}_0$ of
$K^-d$ scattering for the new values of the S--wave scattering lengths
$a^0_0$ and $a^1_0$ obeying the low--energy theorem $a^0_0 + 3\, a^1_0
= 0$. We have shown that the obtained result is
not changed with respect to that calculated in Ref.\cite{IV4}.

For the confirmation of the self--consistency our approach we have
analysed the cross sections for elastic and inelastic channels of
$K^-p$ scattering for laboratory momenta $70\,{\rm MeV}/c \le p_{lab}
\le 150\,{\rm MeV}/c$ of the incident $K^-$--meson. We have shown that
the cross sections for the reactions $K^-p \to K^-p$ and $K^-p \to
Y\pi$, which we have calculated by using the S--wave amplitudes of
elastic and inelastic channels of $K^-p$ scattering at threshold, do
not contradict the experimental data within two standard deviations.
However, the constraints imposed by recent experimental data by the
DEAR Collaboration demand a revision of these data.

The energy level displacement of the ground state of kaonic hydrogen
has been analysed in Ref.\cite{WW04} and Ref.\cite{UM04}. The result
predicted by Borasoy {\it et al.} Ref.\cite{WW04} within the $SU(3)$
chiral effective Lagrangian approach with relativistic coupled
channels technique is equal to 
\begin{eqnarray}\label{label4.1}
- \,\epsilon_{1s} + i\,\frac{\Gamma_{1s}}{2} = 412.13\,f^{K^-p}_0(0) =
-\,235 + \,i\,195\,{\rm eV},
\end{eqnarray}
where $f^{K^-p}_0(0) = -\,0.57 + \,i\,0.47\,{\rm fm}$. It has been
obtained as {\it a result of an ``optimal'' compromise between the
  various existing data sets} Ref.\cite{WW04}. The energy level
displacement of the ground state of kaonic hydrogen, obtained in
Ref.\cite{WW04}, agree with the experimental data by the DEAR
Collaboration within experimental error bars. Our result for the
energy level shift agrees well with that obtained in Ref.\cite{WW04},
whereas the agreement for the values of the energy level width is only
within an accuracy of about $30\,\%$.

The energy level displacement of the ground state of kaonic hydrogen
has been calculated by Mei\ss ner {\it et al.}  Ref.\cite{UM04} under
the assumption of the dominant role of the $\bar{K}^0n$--cusp.  Such a
hypothesis has been proposed by Dalitz and Tuan in 1960
Ref.\cite{RD60} (see also Ref.\cite{RD62}) in the $\mathbb{K}$--matrix
approach in the zero--range approximation. Mei\ss ner {\it et al.}
have argued that the S--wave amplitude
\begin{eqnarray}\label{label4.2}
\tilde{f}^{K^-p}_0(0) =
\frac{\displaystyle \frac{\tilde{a}^0_0 + \tilde{a}^1_0}{2} +
  q_0\,\tilde{a}^0_0\tilde{a}^1_0}{\displaystyle 1 +
  \Big(\frac{\tilde{a}^0_0 + \tilde{a}^1_0}{2}\Big)\,q_0},
\end{eqnarray}
obtained by Dalitz and Tuan within the $\mathbb{K}$--matrix
approach in the zero--range approximation Refs.\cite{RD60,RD62}, can
be derived within a non--relativistic effective Lagrangian approach
based on ChPT by Gasser and Leutwyler Ref.\cite{JG83}.

For the complex S--wave scattering lengths Eq.(\ref{label2.7}) the
energy level displacement of the ground state of kaonic hydrogen,
caused by the $\bar{K}^0n$--cusp, is equal to 
\begin{eqnarray}\label{label4.3}
- \,\epsilon_{1s} +
i\,\frac{\Gamma_{1s}}{2} = 412.13\,f^{K^-p}_0(0) =
 -\,325 +\,i\,248\,{\rm eV}.  
\end{eqnarray}
where $f^{K^-p}_0(0) = -\,0.78 + \,i\,0.60\,{\rm fm}$. This result
agrees well with the experimental data by the KEK Collaboration
Ref.\cite{KEK}
\begin{eqnarray}\label{label4.4}
-\,\epsilon^{(\exp)}_{1s} +
i\,\frac{\Gamma^{(\exp)}_{1s}}{2} = (- 323 \pm 64) + i\,(204\pm
115)\,{\rm eV}.  
\end{eqnarray}
Thus, as has been pointed out by Gasser \cite{JG04}: {\it $\ldots$ the
  theory of $\bar{K}p$ scattering leaves many questions open. More
  precise data will reveal whether present techniques are able to
  describe the complicated situation properly.}

A new set of measurements by the DEAR/SIDDHARTA Collaborations
Ref.\cite{SIDDHARTA}, which is planned on 2006 year and intended for
to reach a precision of the experimental data on the energy level
displacement of the ground state of kaonic hydrogen and kaonic
deuterium at the ${\rm eV}$ level, should place constraints on
theoretical approaches to the description of strong low--energy
$\bar{K}N$ interactions at threshold.

\section*{Acknowledgement}

We are grateful to Torleif Ericson for reading the manuscript and
helpful comments on the results obtained in the paper. The remarks and
comments by Wolfram Weise are greatly appreciated.

\begin{figure}[h]
\psfrag{0}{$0$}\psfrag{20}{$20$}\psfrag{40}{$40$}
\psfrag{60}{$60$}\psfrag{80}{$80$}\psfrag{100}{$100$}
\psfrag{plab}{\hspace{-4mm}$p_{\mathrm{lab}}/\mathrm{MeV}$}
\psfrag{s/mb}{$\sigma/\mathrm{mb}$}
\psfrag{sigma1}{$\sigma_{\Sigma^-\pi+}$}
\psfrag{sigma2}{$\sigma_{\Sigma^+\pi-}$}
\psfrag{sigma3}{$\sigma_{\Sigma^0\pi0}$}
\psfrag{sigma4}{$\sigma_{\Lambda^0\pi0}$}
\centering
\includegraphics[scale=0.8]{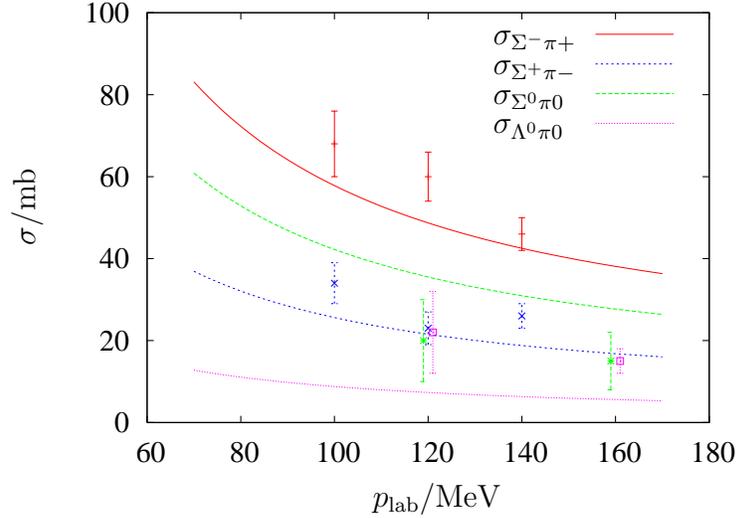}
\caption{Cross--sections for the inelastic reactions $K^-p \to Y\pi$,
where $Y\pi = \Sigma^{-}\pi^+,\Sigma^{+}\pi^-,\Sigma^{0}\pi^0$ and
$\Lambda^0\pi^0$. }
\label{rho-Verteilung}
\end{figure}

\begin{table}[h]
\begin{tabular}{|l||r|r|r|r|r|r|r|r|r|r|r|}
 \hline
$p_{lab}$& 70 & 80 & 90 & 100 & 110 & 120 & 130 & 140 & 150 & 160 & 170\\[0.5ex]
 \hline
$\sigma_{\Sigma^-\pi^+}$ &83.1&72.3&64.1&57.8&52.8&48.7&45.3&42.5&40.1&38.1&36.3\\
 \hline
$\sigma_{\Sigma^+\pi^-}$ &36.9&32.1&28.4&25.6&23.4&21.6&20.0&18.8&17.7&16.8&16.0\\
 \hline
$\sigma_{\Sigma^0\pi^0}$ &60.9&52.9&46.9&42.2&38.5&35.5&33.0&31.0&29.2&27.7&26.4\\
 \hline
$\sigma_{\Lambda^0\pi^0}$&12.8&11.1& 9.8& 8.8& 8.0& 7.3& 6.8& 6.3& 5.9& 5.6& 5.3\\
 \hline
\end{tabular}
\caption{Theoretical values of cross sections for
    inelastic channels of $K^-p$ scattering. The laboratory momentum
    of the incident $K^-$--meson is measured in ${\rm MeV}/c$ and the
    cross sections in ${\rm m b}$.}
\end{table}

\begin{table}[h]
\begin{tabular}{|l||r|r|r|}
 \hline
$p_{lab}$& $90 - 110$ & $110 - 130$& $130 -  150$ \\[0.5ex]
 \hline
$\sigma_{\Sigma^-\pi^+}$& $68 \pm 8$ & $60 \pm 6$ & $46 \pm 4$\\
 \hline
$\sigma_{\Sigma^+\pi^-}$& $34 \pm 5$ & $23 \pm 4$ & $26 \pm 3$\\
\hline
\end{tabular}
\caption{Experimental data on  the cross sections for
  the reactions $K^-p \to \Sigma^-\pi^+$ and $K^-p \to \Sigma^+\pi^-$.
  The laboratory momentum of  the incident $K^-$--meson is measured in 
  ${\rm MeV}/c$ and the cross sections in 
  ${\rm m b}$.}
\end{table}

\begin{table}[h]
\begin{tabular}{|l||r|r|}
 \hline
$p_{lab}$ & 120 & 160 \\[0.5ex]
 \hline
$\sigma_{\Sigma^0\pi^0}$& $ 20 \pm 10$ & $15 \pm 7$\\
 \hline
$\sigma_{\Lambda^0\pi^0}$& $ 22 \pm 10$ & $15 \pm 3$\\
\hline
\end {tabular}
\caption{Experimental data on  the cross sections for
  the reactions $K^-p \to \Sigma^0\pi^0$ and $K^-p \to \Lambda^0\pi^0$.
  The laboratory momentum of  the incident $K^-$--meson is measured 
  in ${\rm MeV}/c$ and the cross sections in  
  ${\rm m b}$.}
\end{table}

\end{document}